# Adapter-dependent Adapter Methylation Assay

Jia Zhang#1, Peng Qi#1, Li Xiao#2, Mengxi Yuan3, Jun Chuan3, Yaling Zeng1, Li-mei Lin1, Yue Gu 4, Yan Zhang 4, Duan-fang Liao*1,5, Kai Li*1,5

#These authors contributed equally to this work
*Correspondence: Kai Li, kaili34@yahoo.com；Duan-fang Liao, dfliao@hnucm.edu.cn

1. National Engineering Research Center for Individualized Diagnosis & Treatment, Hunan University of Chinese Medicine, Changsha, 410208, Hunan, PR China
2. The Second Affiliated Hospital of Soochow University, Suzhou, 215004, PR China
3. Genetalks Co., Changsha, 410208, Hunan, PR China
4. Faculty of Life Sciences and Medicine, Harbin Institute of Technology, Harbin 150001, PR China
5. Epigenetic laboratory, Changsha Health College, Liuyang economic development zone, Changsha, Hunan 410329, PR China

**Abstract:** Sensitive and reliable methylation assay is important for oncogentic studies and clinical applications. Here, a new methylation assay was developed by the use of adapter-dependent adapter in library preparation. This new assay avoids the use of bisulfite and provides a simple and highly sensitive scanning of methylation spectra of circulating free DNA and genomic DNA.

DNA methylation is involved in the development and progression of cancer [1-3]. An increasing number of published papers has documented that methylation assay of circulating free DNA is valuable in early screening of cancer [4]. Recently, methylation-dependent endonucleases mediated next generation sequencing has been reported [5-8], which requires mechanically separation and extraction of 32-bp fragments from enzymatic digestion and limited to decode the palindromic methylated CpG sites in genomic DNA. To simplify the procedure and maximize the decoding ability of methylation-dependent endonucleases, an adapter-dependent adapter methylation assay (ADAMA) is developed. As illustrated in Figure 1, methylated CpGs guide the addition of methylation-dependent adapters through a serial coupling of enzymatic digestion and ligation. Within the methylation-dependent adapter, a particular restriction enzyme site is purposely harbored that can have a remote digestion upstream of the adapter for producing termini that are subjected to the ligation of methylation-independent adapters.

To test if ADAMA works, a chemically synthesized template was used with an adapter having perfectly matched sticky terminal as shown in Figure 2 (panel a). Two restriction endonucleases, BpuEI and MmeI, were respectively harbored in the methylation dependent adapters. As expected, adapter dependent adapter methylation assay functions well (Sequencing chromatographs shown in Figure 2 panel b). The artificial template was then applied with an adapter with a 5' protruding NNNN to address the potential in genomic methylation sequencing for gDNA and cfDNA, which was confirmed by Sanger's sequencing data. In addition to BpuEI and MmeI that can be harbored within the methylation dependent adapter, EcoP15I should be another or even better option as its digestion site and recognition site are separated by 27 nucleotides, which could make bioinformatics analysis more easier for ADAMA data.

Before the development of ADAMA, ideal technologies yearned for years for surveillance the screening and diagnosis of cancer at its early stage are those that they can efficiently introduce NGS adapters flunking the methylated CpGs for scanning methylation spectra for circulating free DNA [9-11]. Methylation dependent endonucleases were applied in genomic methylation sequencing for the palindromic methylated CpG sites [5-8], but some factors restrained its used in cfDNA. First, T-adapters cannot be directly used in sequencing the products digested by methylation-dependent endonucleases from gDNA or from cfDNA. Secondly, as cfDNA is consisted of fragmented DNA with various length, no 32-nt band could be obtained from gel electrophoresis. ADAMA has overcome these disadvantages of conventional assays using the methylation-dependent endonucleases since it covers both palindromic and non-palindromic sites of the particular methylation-dependent endonuclease and overcomes the low efficiency inherited from T-A adapter in gDNA methylation sequencing. More importantly, ADAMA, as purposely designed, provides a practically useful technology for cfDNA methylation assay with sensitivity higher than those that with bisulfite transformation.

The strategy of adapter-dependent adapter does not only replace the sequence-specific primers functionally, but it also massively increased the scalability of ADAMA with no limitation when applied to NGS. A number of unique feature remarks ADAMA one of the ideal technologies for methylation assay. Its sensitivity is greatly higher than all assays with the use of sodium bisulfite by eliminating nucleic acid destruction from BS transformation, which differentiates ADAMA from other technologies in scanning methylation spectra of cfDNA and single cells. Since ADAMA enriched and only assayed the fragments residing methylated CpGs, the timing and cost are more economic than conventional genome-wide methylation sequencing. As shown in Figure 3, methylation dependent enzymes cover over 28 million CpG sites in over 98% of the genome, and may actually cover 100% of the CpGs. All aforementioned advantages of ADAMA, accurate decoding methylated CpGs, enrichment of templates with methylated CpGs, efficiently sequencing templates (ideal efficiency is elimination of library amplification before NGS), and high percentage coverage of CpGs in the genome, allowing surveillance screening and diagnosis of cancer at its early stage be more practical.

As a new method, its precious potential of ADAMA needs to be confirmed by sequencing a large number of basic research samples and clinical samples. At technical point of view, some limitations of ADAMA should be mentioned. First, the direct decoding sequencing surrounds methylated CpG sites is relatively short, between 18-27 bases. Secondly, a whole coverage of CpGs requires the application of more than one methylation dependent endonucleases. Thirdly, some situations may not need to cover tens of millions CpG, instead of much less CpG sites are to be analyzed. In practical application, the aforementioned technical limitations can be well overcome. The relatively short sequences decoded can be rescued by bioinformatic analysis based on the superimposed downstream sequences of the methylated CpG sites. Combination of more than one enzymes is routinely used appreciated the availability of cutsmart buffer from New England Biolab. In situations when less CpG sites to be covered, the scalability of ADAMA can be adjusted by choosing the methylation dependent adapter with different sequencing combinations at its 5' sticking termni, such as from AAAA, CCCC, to NNNN. For example, a terminal with AAAA methylation dependent adapter together with FspEI only cocver 400K CpG sites.

In conclusion, ADAMA is a new methylation assay with a combination of methylation-dependent and independent endonucleases restriction digestions together with their respective adapters for NGS. This new assay integrated the methylation identification ability into NGS platform, makes it an efficient technology in scanning methylation spectra of cfDNA and circulating tumor cells in clinical applications. It is also a powerful technology in fundamental epigenetic studies including methylation analysis of single cells.

**Figure legends**

**Figure 1.** Methylation NGS library is prepared with an upstream non-specific adapter and downstream methylation specific adapter flanking the methylated CG, CHG and CHH sites. The upstream adapters that cohesive to the termini yielded by enzymes other than methylation-dependent endonucleases. The downstream adapters are designed to be ligated to the termini produced by methylation-dependent endonucleases.

**Figure 2.** ADAMA tested with chemically synthesized template. Panel a illustrated the strategy of two different enzymes harbored in methylation dependent adapter. Panel b showed the sequencing chromatographs from the experiments.

**Figure 3.** Illustration showing the coverage of CpGs by ADAMA when three methylation dependent endonucleases are conbined. Theoretically，only two strings of ATCGAY and GTCGAY are not covered by these three enzymes, and so far there no methylation reported from publically available database.

Figure 1.

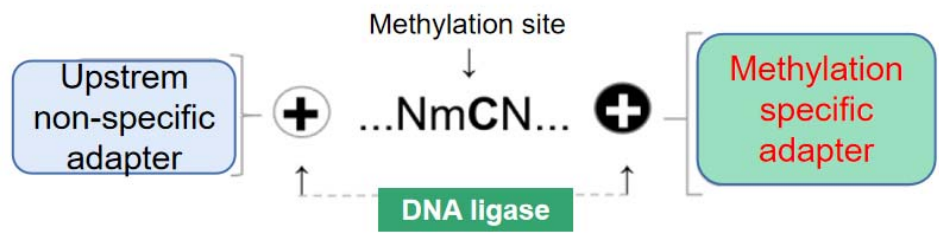

Figure 2.

panel a

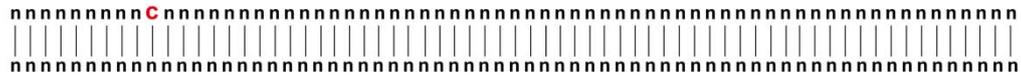

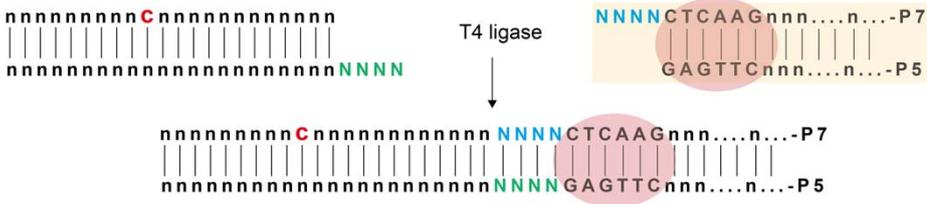

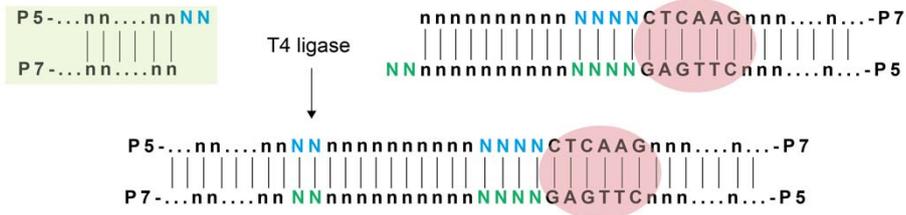

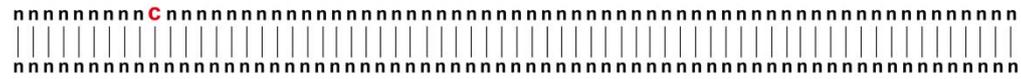

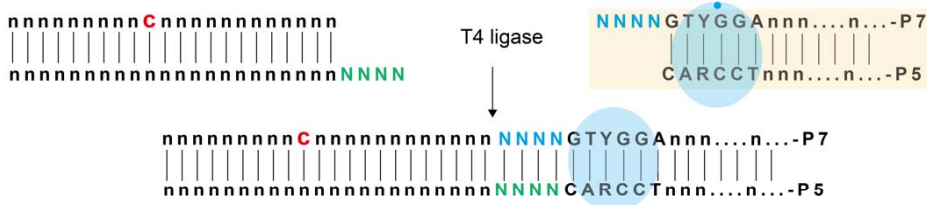

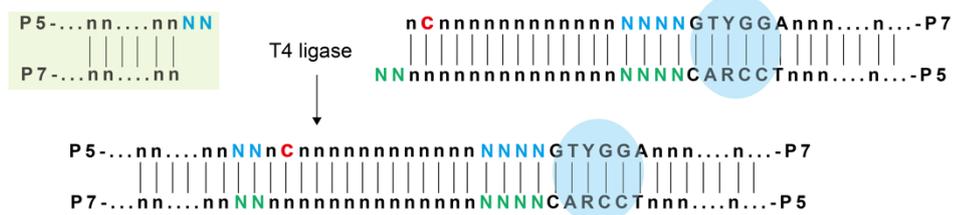

panel b

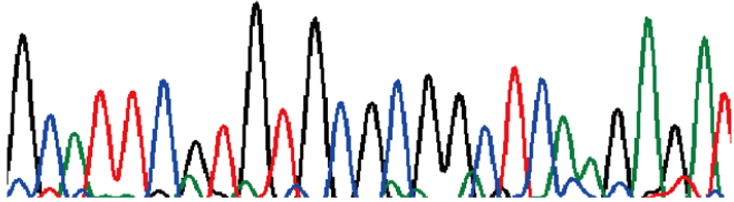

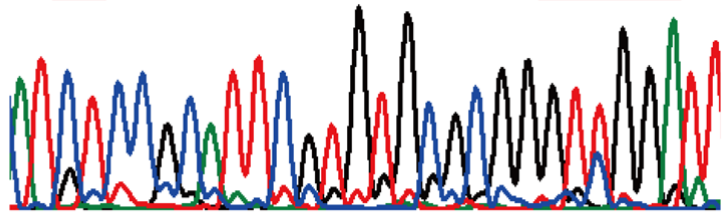

Figure 3.

| Enzymes | | recognition sites | uncovered |
|---|---|---|---|
| McrBC & FspEI : ACG, GCG & CCG | sense | ACG / CCG / GCG / TCG | 1/4 |
| | antisence | TCGA / TCGC / TCGG / TCGT | 1/16 |
| MspJI : CGNA & CGNG | sense | TCGAA / TCGAC / TCGAG / TCGAT | 1/32 |
| | antisence | ATCGAY / CTCGAY / GTCGAY / TTCGAY | 1/64 |